\newcommand{\bfl}{\begin{flushleft}}
\newcommand{\efl}{\end{flushleft}}
\newcommand{\bc}{\begin{center}}
\newcommand{\ec}{\end{center}}

\newcommand\eg {{\it e.g.}}
\newcommand\etc{{\it etc. }}
\newcommand\cf {{\it cf.  }}

\def\be{\begin{eqnarray}}
\def\ee{\end{eqnarray}}
\def\pp{+\hspace{-0.04in}+}

\newenvironment{draftequation}[1]{\be\label{#1}}{\ee}
\newcommand\lag{{\sf L}}
\newcommand\qqq{{\cal Q}}
\newcommand\bbe[1]{\begin{draftequation}{#1}}
\newcommand\eee{\end{draftequation}}

\def\half{{\textstyle{1 \over 2}}}

\def\del{\Delta}
\def\nab{\nabla}

\documentstyle[12pt]{article}
\begin{document}
\begin{titlepage}
\begin{flushright}
ITP-SB-93-34\\
USITP-93-15\\
OSLO-TP 1-94\\
hep-th/9401091
\end{flushright}
\bigskip
\Large
\begin{center}
\bf{New $N=4$ superfields and $\sigma$-models}\\

\bigskip

\normalsize
by\\
\bigskip

U.\ Lindstr\"om\footnote{email:ul@vana.physto.se} \\
{\it Institute of Theoretical Physics\\
University of Stockholm\\
Box 6730\\
S-113 85 Stockholm
SWEDEN}\\
and\\
{\it Department of Physics\\
P.\ O.\ Box 1048 Blindern\\
University of Oslo\\
N-0316 NORWAY}\\

\bigskip

\bigskip

I.T.Ivanov\footnote{email:iti@max.physics.sunysb.edu}
and M.\ Ro\v cek\footnote{email:rocek@insti.physics.sunysb.edu} \\
{\it Institute for Theoretical Physics\\
State University of New York\\
 Stony Brook, NY 11794-3840
USA\\}

\end{center}
\vspace{1.0cm}
\normalsize
{\bf Abstract:} In this note, we construct new
representations of $D=2,\ N=4$ supersymmetry which do not involve
chiral or twisted chiral multiplets.  These multiplets may make it
possible to circumvent no-go theorems about $N=4$ superspace formulations
of WZWN-models.
\end{titlepage}
\eject
\setcounter{footnote}{0}

\begin{flushleft}
\section{Introduction}
\end{flushleft}

In this note, we introduce some new representations of $N=4$
supersymmetry in $D=2$. These are characterized by having no $N=2$
component superfields that are chiral or twisted chiral.
Such multiplets are important, because in the $N=4$ theory, due
to arguments based on the dimension of the superspace action (see for
example \cite{kahler,GHR}), a good deal of the dynamics of the theory
depends simply on the choice of multiplets.  Indeed, previously known
multiplets cannot be used to describe generic $N=2,4$ WZW-models
\cite{schout},\cite{bbkim}.

\begin{flushleft}
\section {New Representations of $N=4$ Supersymmetry}
\end{flushleft}

\bigskip
As may be seen from a dimensional analysis of the superspace
measure and the superfield component content, to construct
superspace actions for higher $N$ in $D=2$ one needs to find
invariant subspaces and corresponding restricted measures. Such subspaces are
analogous to $N=1$, $D=4$ chiral and antichiral superspaces. In
a series of papers \cite{GHR,LR,BLR,projective}, to this end we have
constructed and utilized a projective superspace. In the present $N=4$
context it is introduced as follows\footnote{A more detailed discussion
may be found in \cite {LR}.}:
\bigskip

\noindent The complex $SU(2)$ doublet spinor derivatives $D_{a\pm}$, $\bar
D^b_{\pm}$ that describe $N=4$ supersymmetry obey the commutation
relations
\bbe{CREL}
\left\{{D_{a\pm},\bar D^b_{\pm}}\right\}=i\delta ^b_a\partial_{\buildrel
\pp \over =}
\eee
(all others vanish). We will work with  $N=2$ superfields and identify
$D_\pm\equiv
D_{1 \pm}$ as the $N=2$ spinor covariant derivative and $Q_\pm \equiv
D_{2\pm}$ as
the generator of the non-manifest supersymmetries. As described in
\cite{BLR} we use
two complex variables $\zeta$ and $\xi$ to define a set of anticommuting
left and
right derivatives:
\bbe{NABLA}
\nabla_+=D_++\zeta Q_+\ , \qquad & \nabla_-=D_-+\xi Q_-\ , \cr & \cr
\bar\nabla_+=\bar D_+-\zeta^{-1} \bar Q_+\ , \qquad & \bar \nabla_-=
\bar D_--\xi^{-1}\bar Q_-\ .
\eee
A real structure $R$ acts on $\zeta$ and $\xi$ by hermitian conjugation
composed with
the antipodal map, i.e.;
\bbe{REAL}
R\zeta =-\bar\zeta^{-1}, \qquad R\xi =-\bar\xi^{-1}
\eee
Clearly $\nabla_\pm =R\bar\nabla_\pm$, so $R$ preserves the subspaces
defined by the
derivatives (\ref{NABLA}). We then consider superfields $\eta(\zeta ,\xi)$,
specify
the $\zeta$ and $\xi$ dependence (typically as a series expansion) and
require that
$\eta$ is annihilated by the derivatives in (\ref{NABLA}). In general, if
we write
\bbe{ETADEF}
\eta = \sum\limits_{}^{} {\zeta^n\xi^m\eta_{nm}}
\eee
the $N=4$ constraints that $\eta$ is annihilated by the
derivatives in
(\ref{NABLA}) lead to the component relations
\bbe{DETA}
D_+\eta_{nm}+Q_+\eta_{n-1,m}=0,&
\qquad {\bar D}_+\eta_{nm}-{\bar Q}_+\eta_{n+1,m}=0\cr
D_-\eta_{nm}+Q_-\eta_{n,m-1}=0, &\qquad {\bar D}_-\eta_{nm}-
{\bar Q}_-\eta_{n,m+1}=0.
\eee
We may also specify $\eta$
further by, \eg, a reality condition such as
\bbe{RCOND}
\eta =R\bar\eta
\eee

To construct $N=4$ actions for $N=4$ superfields we use a second set of
linearily independent covariant spinor derivatives:
\bbe{DELTA}
\Delta_+=D_+-\zeta Q_+\ ,\qquad &\Delta_-=D_--\xi Q_-\ ,\cr & \cr
\bar\Delta_+=\bar D_++\zeta^{-1}\bar Q_+\ ,\qquad &\bar\Delta_-=\bar
D_-+\xi^{-1}\bar Q_-\ .
\eee
An action may then be written as
\bbe{N4ACT}
S=\frac1{16}\int{d^2\sigma }\int_{C}{d\zeta}\int_{C'}{d\xi}\Delta_+\Delta_
-\bar\Delta_+\bar\Delta_- \, L\left({\eta(\zeta ,\xi );\zeta , \xi }\right)
\eee
where $C$ and $C'$ are some appropriate contours.

Using
\bbe{delnab}
\del_+=2D_+-\nab_+\ ,\qquad &\del_-=2D_--\nab_-\ ,\cr & \cr
\bar\del_+=2\bar D_+-\bar\nab_+\ ,\qquad &\bar\del_-=
2\bar D_--\bar\nab_-\ ,
\eee
and (of course) $\nab_\pm\eta=\bar\nab_\pm\eta=0$, the $N=2$ superspace
form of the
action (\ref{N4ACT}) is:
\bbe{N2ACT}
S=\int{d^2\sigma }D^2\bar D^2\int_{C}{d\zeta}\int_{C'}{d\xi}
L\left({\eta(\zeta ,\xi );\zeta , \xi }\right)
\eee

In \cite{BLR} this general setting was applied to study a particular $\eta$
obeying
\bbe{REAL2}
R\bar\eta = (-\zeta)^{-N}(-\xi)^{-M}\eta ,
\eee
where the sum in (\ref{ETADEF}) was restricted to be from $0$ to $N,M$.
Among the component superfields we found the usual chiral and twisted chiral
superfields, as well as semi-chiral and semi-antichiral fields, i.e. fields
$\phi$
and $\tilde \phi$ that obey {\it only}
\bbe{SEMI}
D_{1+}\phi\equiv D_+\phi =0, \quad D_{1-}\tilde\phi\equiv D_-\tilde\phi =0.
\eee
Here we further extend the set of $N=4$ multiplets to include new
types
that contain {\em no} chiral or twisted chiral $N=2$ component superfields.

We begin with a general multiplet of the type (\ref{ETADEF}).
If we further impose the reality condition (\ref{RCOND}), then the definition
(\ref{REAL}) implies
\bbe{ETAR}
\eta_{nm}=(-)^{n+m}\bar\eta_{-n,-m}.
\eee
So far the expansion in (\ref{ETADEF}) is quite general: $n$ and $m$ range
over the
integer numbers $\bf Z$ from $-\infty$ to $+\infty$, and we have not
worried about
convergence properties, \etc Note that the constraints (\ref{DETA}) are
translation
invariant; this leads us naturally to restrict the expansion by
\bbe{LCYL}
\eta_{n+k,m}=\eta_{nm}
\eee
for some fixed $k\in \bf Z$ (for $\eta$ real in the sense of (\ref{ETAR}),
$k$ must be
even). We shall call (\ref{LCYL}) a left cylindrical constraint. Similarily
\bbe{RCYL}
\eta_{n,m+k}=\eta_{nm}
\eee
is a right cylindrical constraint and
\bbe{TOR}
\eta_{n+k,m+l}=\eta_{nm}, \qquad k,l \in \bf Z
\eee
is called a toroidal constraint.

\bigskip

The components of a real $\eta$ obeying a toroidal constraint will obey the
constraints (\ref{DETA}). We see that this only restricts the components'
transformation
properties under the nonmanifest supersymmetries. A dimensional analysis of
the measure
in the action (\ref{N2ACT}) shows that we need $D_{\pm}$ constraints on the
components
to generate dynamics. The real toroidal $\eta$'s thus correspond to purely
auxiliary
multiplets.

\bigskip

Cylindrical $\eta$'s have some components that are semi-(anti)chiral and may
hence be used to construct dynamical actions. Let us illustrate this in a
particular
case. We choose two real $\eta$'s with components $\eta_{nm}$ and $\chi_{nm}$
respectively. On the first we impose a left cylindrical constraint
(\ref{LCYL}) with
$k=2$ and on the second we impose a right cylindrical constraint
(\ref{RCYL}) with
the same $k$. We also restrict the $m$ and $n$ range so that we have the
following
set of conditions
\bbe{ETACHI}
\eta_{n+2,m}=\eta_{nm}, \qquad \eta_{nm}=0, \quad |m|>1\cr
\chi_{n,m+2}=\chi_{nm}, \qquad \chi_{nm}=0, \quad |n|>1,
\eee
in addition to the constraints (\ref{DETA}) and (\ref{ETAR}) (that hold for
both
$\eta_{nm}$ and $\chi_{nm}$). Explicitly this yields
\bbe{EXPLICIT}
&\eta_{11}=\bar\eta_{1,-1}\ , \quad \eta_{00}=\bar\eta_{00}\ , \quad
\eta_{10}=-\bar\eta_{10}\ , \quad \eta_{01}=-\bar\eta_{0,-1}\ ;\cr &\cr
&D_-\eta_{n,-1}=0\ , \quad \bar D_-\eta_{n,1}=0\ ;\cr &\cr
&\chi_{11}=\bar\chi_{-1,1}\ , \quad \chi_{00}=\bar\chi_{00}\ , \quad
\chi_{01}=-\bar\chi_{01}\ , \quad \chi_{10}=-\bar\chi_{-1,0}\ ;\cr &\cr
&D_+\chi_{-1,m}=0\ , \quad \bar D_+ \chi_{1m}=0\ .
\eee
Note that $\eta _{00}$ and $\chi _{00}$ are real while $\eta _{10}$ and
$\chi _{01}$
are imaginary.  We now discuss what is required from a Lagrangian $L(\eta
,\chi )$ for
the action (\ref{N2ACT}) to be $N=4$ supersymmetric. From
\cite{SCALARTENSOR} we know
that an action constructed out of ordinary $N=2$ (anti)chiral superfields
is $N=4$
supersymmetric if and only if the Lagrangian satisfies a generalized
Laplace equation. Here we expect to find some similar requirement.

Because of the issues of convergence that we have ignored, we cannot
simply use the action (\ref{N4ACT}) or its $N=2$ reduction
(\ref{N2ACT}). We therefore look for an invariant action directly.  A
non-manifest supersymmetry transformation generated by $Q_+$ acting on
$L(\eta ,\chi )$ has the following effect
\bbe{SUSYL}
Q_+L(\eta ,\chi )=-(L_{\eta_{nm}}D_+\eta_{n+1,m}+L_{\chi_{nm}}D_+\chi_{n+1,m})
\eee
where we have used the conditions (\ref{DETA}). For the action to be
invariant we
require $Q_+L$ to be a total (super)derivative, which implies
$D_+Q_+L(\eta ,\chi)=0$.  This leads to
\bbe{DQL}
L_{{\eta_{n+1,m}}{\eta_{ij}}}D_+\eta_{ij}D_+\eta_{nm}+(L_{{\chi_{ij}}{\eta_{
n+1,m}}}
-L_{{\chi_{i-1,j}}{\eta_{nm}}})D_+\chi_{ij}D_+\eta_{nm} &\cr&\cr
+L_{{\chi_{n-1,m}}{\chi_{ij}}}D_+\chi_{ij}D_+\chi_{nm}=0&\ .
\eee
Exploring this equation using the explicit relations (\ref{EXPLICIT}),
we find a set of equations for the derivatives of $L$. To present them
and their solutions it is convenient to introduce the following linear
combinations of the fields:
\bbe{NEWF}
\eta_1=\eta_{11}+\eta_{01}, &\eta_0=\eta_{00}+\eta_{10}, &\eta_2
=\eta_{11}-\eta_{01} \\
\chi_1=\chi_{11}+\chi_{10}, &\chi_0=\chi_{00}+\chi_{01}, &\chi_2
=\chi_{11}-\chi_{10} \ .
\eee
Vanishing of the mixed derivatives term in (\ref{DQL}) implies the
following set of equations:
\bbe{MIXED}
&\begin{array}{llllllllll}
L_{\chi_0 \eta_0} &= &L_{\chi_1 \eta_0} &= &L_{\bar\chi_2 \eta_0} \; , \;\;\;\;
&L_{\bar\chi_0 \eta_0} &= -&L_{\bar\chi_1 \eta_0} &= -&L_{\chi_2
\eta_0}\; ,\\
L_{\chi_0 \bar\eta_0} &= -&L_{\chi_1 \bar\eta_0} &= -&L_{\bar\chi_2 \bar\eta_0}
\; ,
&L_{\bar\chi_0 \bar\eta_0} &= &L_{\bar\chi_1 \bar\eta_0} &= &L_{\chi_2
\bar\eta_0}\; ,\\
L_{\chi_0 \eta_1} &= &L_{\chi_1 \eta_1} &= &L_{\bar\chi_2 \eta_1} \; ,
&L_{\bar\chi_0 \eta_1} &= - &L_{\bar\chi_1 \eta_1} &= - &L_{\chi_2
\eta_1} \; ,\\
L_{\chi_0 \bar\eta_1} &= - &L_{\chi_1 \bar\eta_1} &= - &L_{\bar\chi_2
\bar\eta_1} \; ,
&L_{\bar\chi_0 \bar\eta_1} &= &L_{\bar\chi_1 \bar\eta_1} &= &L_{\chi_2
\bar\eta_1} \; ,\\
L_{\chi_0 \bar\eta_2} &= &L_{\chi_1 \bar\eta_2} &= &L_{\bar\chi_2
\bar\eta_2} \; ,
&L_{\bar\chi_0 \bar\eta_2} &= - &L_{\bar\chi_1 \bar\eta_2} &= - &L_{\chi_2
\bar\eta_2}\; ,\\
L_{\chi_0 \eta_2} &= - &L_{\chi_1 \eta_2} &= - &L_{\bar\chi_2
\eta_2} \; ,
&L_{\bar\chi_0 \eta_2} &= &L_{\bar\chi_1 \eta_2} &= &L_{\chi_2
\eta_2} \; .
\end{array}&\cr
&{}&
\eee
The pure $\eta$ derivatives yield
\bbe{ETADER}
L_{\eta_0 \bar\eta_0} = L_{\eta_0 \bar\eta_1} = L_{\eta_0 \eta_2} = 0
\; , \nonumber \\
L_{\eta_1 \bar\eta_0} = L_{\eta_1 \bar\eta_1} = L_{\eta_1 \eta_2} = 0
\; , \nonumber \\
L_{\bar\eta_2 \bar\eta_0} = L_{\bar\eta_2 \bar\eta_1} = L_{\eta_2
\bar\eta_2} = 0 \; .
\eee
and pure $\chi$ derivatives
\bbe{CHIDER}
L_{\chi_0 \chi_0} = L_{\chi_1 \bar\chi_2}\ ,
\qquad & L_{\bar\chi_0 \bar\chi_0} =
L_{\chi_2 \bar\chi_1} \ , \nonumber \\
L_{\chi_0 \bar\chi_1} = - L_{\bar\chi_0 \bar\chi_2} \ , \qquad &
L_{\chi_0 \chi_2} = - L_{\chi_1 \bar\chi_0} \ , \nonumber
\eee
\bbe{CHIDER2}
L_{\chi_0 \bar\chi_0} = - L_{\chi_1 \bar\chi_1} = - L_{\chi_2 \bar\chi_2} \ .
\eee
There are also the relations that result from $D_-Q_-L = 0$. They can be
obtained from the above via the substitution $\eta \leftrightarrow \chi$ (not
all equations from the full set are independent; in particular, the
complex conjugate relations have already been included).

 A solution for the supersymmetric Lagrangian $L$ is a
linear combination of the following fuctions $L_1, L_2, L_3, L_4 $ (and the
corresponding complex conjugate expressions):

\bbe{LETA}
L_1(\eta)=\int d\xi \,
\lag_1\Big(\xi\eta_1+\eta_0+
\frac1\xi\bar\eta_2); \xi\Big)\ ,
\eee

\bbe{LCHI}
L_2(\chi)=\int d\zeta \,
\lag_2\Big(\zeta\chi_1+\chi_0+\frac1\zeta\bar\chi_2); \zeta\Big)\ ,
\eee

\bbe{LEC}
L_3(\eta ,\chi)=
L_3\Big(\eta_1+\eta_0+\bar\eta_2, \chi_1+\chi_0+\bar\chi_2\Big)\ ,
\eee

\bbe{REC}
L_4(\eta ,\chi)=
L_4\Big(\eta_1-\eta_0+\bar\eta_2, \bar\chi_1-\bar\chi_0+\chi_2\Big)\ .
\eee

Since the fields $\vec y_i = ( \chi_0, \bar\chi_0, \eta_0, \bar\eta_0)$ are
unconstrained, their equations of motion will imply:
\bbe{CONS}
L_{\chi_0} = L_{\bar\chi_0} = L_{\eta_0} = L_{\bar\eta_0} = 0
\eee
As long as $L_{y_iy_j}$ is nondegenerate we can solve these for $y_i$
in terms of the remaining fields $\vec x= $ $ (\vec\chi, \vec\eta) $
\bbe{VECT}
\vec\chi_i = (\chi_1, \bar\chi_1, \chi_2, \bar\chi_2) \ , \ \
\vec\eta_i = (\eta_1, \bar\eta_1, \eta_2, \bar\eta_2)
\eee
and replace $y_i$ in the Lagrangian.

To interpret these results, we start from the $N=2$ superspace reduction
and continue down to
$N=1$ superspace in the standard way. $N=1$ superspace derivatives $\nabla$
and extra supersymmetry generators $\cal Q$ are defined as
\bbe{N1DS}
\nabla \equiv D+\bar D\ ,\qquad {\cal Q} \equiv i(D-\bar D)\ ,
\eee
 Starting from (\ref{N2ACT}) we obtain
\bbe{N1ACT}
S = \int {d^2\sigma}  \nabla_+\nabla_-{\cal Q}_+{\cal Q}_- \,\,
L(x_i,y_j(x_i))\qquad\qquad\qquad\qquad\qquad\nonumber \\
 = \int {d^2\sigma}  \nabla_+\nabla_- \Big( M_{x_ix_j}(x_k) \; {\cal
Q}_+x_i \;{\cal Q}_-x_j + L_{x_i} \; \qqq_+\qqq_-x_i \Big) \ .
\eee
where
\bbe{FUNC}
M_{x_ix_j} (x_k) = L_{x_ix_j} - L_{x_iy_k}\Big( L_{y_ly_k}\Big)^{-1} L_{y_lx_j}
\eee
We now use  the constraints (\ref{EXPLICIT}) to eliminate the extra
supersymmetry generators ${\cal Q}$ in terms of $N=1$
$\nabla$-derivatives wherever possible and integrate
by parts to rewrite the action as
\bbe{NEWACT}
S = \int d^2\sigma  \nabla_+ \nabla_- \; \big( \nabla_+\vec{\chi_i} \; A_{ij}
\;
\nabla_-
\vec{\eta_j} + \nabla_+\vec{\chi_i} \; B_{ij} \; {\cal Q}_-\vec{\chi_j} +
\nabla_+
\vec{\eta_i} \; C_{ij} \; {\cal Q}_-\vec{\chi_j} \nonumber \\ + {\cal
Q}_+\vec{\eta_i} \; D_{ij} \;
 \nabla_-\vec{\eta_j} + {\cal Q}_+\vec{\eta_i} \; E_{ij} \;
\nabla_-\vec{\chi_j} +
{\cal Q}_+\vec{\eta_i} \; F_{ij} \; {\cal Q}_-\vec{\chi_j} \big) \;\;\;\;\
\eee
where $A,B,C,D,E,F$ are $4\times 4$ matrices:
\bbe{MATA}
A_{ij} = (IMI)_{\chi_i \eta_j} \ , \qquad B_{ij} =
[I,M]_{\chi_i\chi_j} \ ,
\nonumber
\eee
\bbe{MATC}
C_{ij} = -(MI)_{\eta_i \chi_j} \ , \qquad
D_{ij} = -[I,M]_{\eta_i \eta_j} \ , \nonumber
\eee
\bbe{MATE}
 E_{ij} = -(IM)_{\eta_i \chi_j} \ ,  \qquad
F_{ij} = M_{\eta_i \chi_j}\ ,
\eee
and
\bbe{DEFJ}
I = \left( \begin{array}{cc} J & 0\\ 0 & J
\end{array} \right)\ , \qquad J=\left( \begin{array}{ccccc} i & 0
&0&0\\ 0 & -i&0&0\\0&0&i&0\\0&0&0&-i\end{array} \right)\ .
\eee
The fields with explicit $\cal Q$ generators remaining are auxiliary
spinors, and may be eliminated by completing the square to obtain:
\bbe{RACT}
S = \int d^2\sigma  \nabla_+ \nabla_- ( \nabla_+\vec\chi_i A_{ij}
\nabla_-\vec\eta_j \  \hspace{2in} \nonumber \\
\ - (\nabla_+ \vec\chi_i B_{ik} +
\nabla_+\vec\eta_i C_{ik})
(F_{lk})^{-1} (D_{lj} \nabla_-\vec\eta_j + E_{lj} \nabla_-\vec\chi_j) )
\eee
The matrix $F_{ij}$ is invertible as long as
\bbe{INVR1}
 L_{\eta_0\eta_1} \neq L_{\eta_0\bar\eta_2}  \ ,\nonumber
\eee
\bbe{INVR2}
 L_{\chi_0\chi_1} \neq L_{\chi_0\bar\chi_2} \ ,\nonumber
\eee
\bbe{INV3}
 L_{\eta_0 \chi_0}\neq0 \;\;\ ,\qquad   L_{\eta_0 \bar\chi_0} \neq 0 \ ,
\eee
(and their complex conjugates) are satisfied \footnote{This relations are not
obvious, and were found using the algebraic manipulation program Maple.}.
The explicit expressions for the action
(\ref{LETA}),(\ref{LCHI}),(\ref{LEC}) and (\ref{REC}) show that these
conditions are generically satisfied by our Lagrangian.
We can write the final $N=1$ superspace action in the form
\bbe{FINACT}
 S = \int d^2\sigma  \nabla_+\nabla_- \; \left( \nabla_+x_i \; T_{ij}(x_k) \;
\nabla_-x_j \right) \ ,
\eee
where the matrix $T_{ij}$ has $4\times 4$ blocks given by:
\bbe{MATTT}
T = \left( \begin{array}{cc} -BF^{-1}E & A-BF^{-1}D \\ -CF^{-1}E &
-CF^{-1}D \end{array} \right)
\eee
The extra supersymmetry generators $\cal{Q}$ are given by:
\bbe{QPLUS}
{\cal{Q}}_+ \vec x = \left( \begin{array}{cc} J & 0 \\ -{F^{-1}}^{t} B^{t} &
-{F^{-1}}^{t} C^{t} \end{array} \right) \nabla_+ \vec x
\eee
\bbe{QMIN}
{\cal{Q}}_- \vec x = \left( \begin{array}{cc} -F^{-1} E & -F^{-1} D \\
0 & J \end{array} \right) \nabla_- \vec x
\eee
Here we have used the vector notation from (\ref{VECT}) once again,
and $J$ is defined in (\ref{DEFJ}). The
expressions for the remaining supersymmetry generators $Q_{\pm}$
and $\bar Q_{\pm}$ may be obtained from (\ref{DETA}),(\ref{N1DS}) and
(\ref{QPLUS}),(\ref{QMIN}):
\bbe{Q+}
Q_+\vec \eta &=& \left( \begin{array}{ccccc} -1 & 0 &0&0\\ 0 & 1&0&0\\
0&0&1&0\\0&0&0&-1\end{array} \right)D_+\vec \eta, \cr
&{\ }&\cr
&{\ }& \cr
Q_+\vec \chi &=&\left( \begin{array}{ccccc} 0 & 0 &0&0\\ 0 & -1&0&0\\
0&0&0&0\\1&0&0&0\end{array} \right)
\left(L_{yy}\right)^{-1}\left(L_{yx}\right)D_+\vec x \ ,
\eee
\bbe{Q-}
Q_-\vec \chi &=& \left( \begin{array}{ccccc} -1 & 0 &0&0\\ 0
&1&0&0\\0&0&1&0\\0&0&0&-1\end{array} \right)D_-\vec \chi, \cr
&{\ }& \cr
&{\ }& \cr
Q_-\vec \eta &=&\left( \begin{array}{ccccc} 0 & 0 &0&0\\ 0 & 0&0&-1\\
0&0&0&0\\0&0&1&0\end{array} \right)\left(L_{yy}\right)^{-1}
\left(L_{yx}\right)D_-\vec x \ ,
\eee
where
$\left(L_{yx}\right)$ is the matrix with entries $L_{y_ix_j}$, \etc ,
and $D=\half (\nabla - i{\cal Q})$.
In $N=1$ superspace, extra supersymmetries correspond to complex
structures; we may read off the left and
right complex structures $J^A_\pm$ from
${\cal Q}^A_\pm \vec x\equiv J^A_\pm \nabla_\pm\vec x$ (with ${\cal Q}^A =
\{{\cal Q}, (Q+\bar Q),  i(Q-\bar Q)\}$) \cite{GHR}.

Other $N=4$ multiplets of this type can be constructed by modifying
the $\zeta$ and $\xi$ dependence.  In particular, one can consider a
{\it complex} multiplet with periodicity $k=1$ (\cf (\ref{LCYL})); a quick
analysis shows no interesting new features.

\begin{flushleft}
\section{Conclusions}
\end{flushleft}

We have found a new class of $N=4$ multiplets that have no chiral or
twisted chiral $N=2$ component superfields. It is known that general
WZWN-models cannot be described in extended superspace in terms of
{\em only} chiral and twisted chiral superfields \cite{schout}; it was
hoped that multiplets, introduced in \cite{BLR}, with semi-chiral and
semi-antichiral superfields as well as chiral and twisted chiral
superfield could be used.  Recently, we have shown that in many cases
the description cannot involve {\em any} chiral or twisted chiral
superfields \cite{bbkim}; the new multiplets introduced here are thus the only
known candidates for describing WZWN-models in $N=4$ superspace.

\bigskip
\begin{flushleft}
{\bf Acknowledgments}

It is a pleasure to thank the ITPs at Stony Brook and Stockholm,
as well as the Physics Department at Oslo University, for hospitality.
UL acknowledges partial support from the NFR under Grant No.\ F-FU
4038-300 and NorfA under Grant No.\ 93.35.088/00, and MR acknowledges
partial support from the NFS under Grant No.\
PHY 93 09888.
\end{flushleft}

\eject


\begin{thebibliography}{99}

\bibitem{kahler}
B.\ Zumino,
\newblock {\it Phys.\ Lett.}, {\bf 87B} (1979) 203:
 {L.\ Alverez-Gaum\' e} and D.\ Z.\ Freedman
\newblock{\it Commun.\ Math.\ Phys.}, {\bf 80} (1981) 443:

\bibitem{GHR}
{S.\ J.\ Gates,} {C.\ M.\ Hull} and M.\ Ro\v cek,
\newblock {\it Nucl.\ Phys.}, {\bf B248} (1984) 157.

\bibitem{schout}
{M.\ Ro\v cek,} {K.\ J.\ Schoutens,} and A.\ Sevrin.,
\newblock{\it Phys.\ Lett.}, {\bf B265} (1991) 303.

\bibitem{bbkim}
{B.\ B.\ Kim,} and M.\ Ro\v cek.,
\newblock {In preparation}.

\bibitem{LR}
{U.\ Lindstr\"om,} and M.\ Ro\v cek.,
\newblock {\it Commun.\ Math.\ Phys.}, {\bf 115} (1988) 21.

\bibitem{BLR}
{T.\ Buscher,} {U.\ Lindstr\"om,} and M.\ Ro\v cek.,
\newblock {\it Phys.\ Lett.}, {\bf 202B} (1988) 94.

\bibitem{SCALARTENSOR}
{U.\ Lindstr\"om,} and M.\ Ro\v cek.,
\newblock {\it Nucl.\ Phys.}, {\bf B222} (1983) 309.

\bibitem{projective}
{A. Karlhede,} {U.\ Lindstr\"om,} and M.\ Ro\v cek.,
\newblock {\it Phys.\ Lett.}, {\bf 147B} (1984) 297;\\
{J. Grundberg} and U. Lindstr\"om.,
\newblock {\it Class.\ Quantum Grav.}, {\bf 2} (1985) L33;\\
{P.\ Howe,} {A.\ Karlhede,} {U.\ Lindstr\"om,} and M.\ Ro\v cek.,
\newblock{\it Commun.\ Math.\ Phys.}, {\bf 108} (1987) 535;\\
{U.\ Lindstr\"om,} and M.\ Ro\v cek.,
\newblock{\it Commun.\ Math.\ Phys.}, {\bf 128} (1990) 191.



\end{thebibliography}
 \end{document}